\documentclass[a4paper, conference]{IEEEtran}

\usepackage{amsmath,amsfonts,amssymb,latexsym,amsthm,verbatim}

 \theoremstyle{plain}
  \newtheorem{theorem}{Theorem}[section]
  \newtheorem{corollary}[theorem]{Corollary}
  
  \newtheorem{lemma}[theorem]{Lemma}  
  \newtheorem{definition}[theorem]{Definition}
  
  \newtheorem{remark}[theorem]{Remark}
 \newtheorem{example}[theorem]{Example}

\newcommand{\ep}{{\mathbb E}}

\newcommand{\ZZ}{{\mathbb Z}}
\newcommand{\CC}{{\mathbb C}}

\newcommand{\Id}{{\mathbf I}}

\newcommand{\vc}[1]{{\mathbf{#1}}}

\newcommand{\nadd}[1]{\boxplus_{#1}}
\newcommand{\TT}{{\mathcal{T}}}
\newcommand{\LL}{{\mathcal{L}}}
\newcommand{\com}[1]{\vc{#1}_c}
\newcommand{\wth}{\widetilde{H}}
\newcommand{\wtp}{\widetilde{\phi}}

\newcommand{\Lap}[1]{\left[L  #1 \right]}
\newcommand{\Lapinv}[1]{\left[L^{-1}  #1 \right]}

\title{A de Bruijn identity for discrete random  variables}
\author{
\IEEEauthorblockN{Oliver Johnson}
\IEEEauthorblockA{School of Mathematics\\
University of Bristol\\
Bristol, BS8 1TW, UK \\
Email: maotj@bristol.ac.uk}
\and \IEEEauthorblockN{Saikat Guha}
\IEEEauthorblockA{Quantum Information Processing group\\
Raytheon BBN Technologies\\
Cambridge, MA 02138, USA \\
Email: saikat.guha@raytheon.com}
}
\date{\today}

\begin{document}

\maketitle

\begin{abstract}
We discuss properties of the ``beamsplitter addition" operation, which provides a non-standard scaled convolution of random variables
supported on the non-negative integers. We give a simple expression for the action of  beamsplitter addition using
 generating functions. We use this to give a self-contained and purely classical proof of 
a heat equation and de Bruijn identity, satisfied when one of the variables is geometric.
\end{abstract}

\section{Introduction and notation}

Stam \cite{stam} showed that addition of  independent continuous random variables satisfies the de Bruijn identity
\cite[Eq. (2.12)]{stam}, in that the derivative of  entropy under the addition of a normal is Fisher Information.  This  identity
 lies at the heart of many analyses of entropy under addition, including Stam's proof of Shannon's Entropy Power Inequality (EPI)
\cite{shannon2} and Barron's information theoretic Central Limit Theorem \cite{barron}, and Madiman and Barron's proof
of monotonicity of entropy under addition of independent identically-distributed (i.i.d.) continuous random variables~\cite{madiman}. 
These results  have the Gaussian distribution at their heart, relating to the Gaussian maximum entropy property and
closure of the Gaussian family under addition (``Gaussian $+$ Gaussian $=$
Gaussian''). The de Bruijn identity follows because 
 the densities in question satisfy the heat equation \cite[Eq. (5.1)]{stam}.

There have been many attempts to develop a corresponding theory for discrete random variables, often focussing
on the  Poisson family  which  is closed under standard integer addition
(``Poisson $+$ Poisson $=$ Poisson''). Results
 in this context include Poisson limit theorems \cite{johnson11, johnson26}, maximum entropy property \cite{johnson21} and
 monotonicity result \cite{johnson27}. However, \cite{johnson21} and \cite{johnson27} rely on the assumption
of ultra-log-concavity (ULC), meaning that they are more restrictive than their Gaussian counterparts.

In this paper we prove new properties of what we refer to as the `beamsplitter addition'  $\nadd{\eta}$ 
(see Definition \ref{def:beamsplit}) of  random variables
supported on the non-negative integers $\ZZ_+$. By design, the geometric family is closed under the action of $\nadd{\eta}$ (``Geometric $\nadd{\eta}$ Geometric $=$ Geometric'').
Geometric is more natural than the Poisson since no auxiliary assumptions such as ULC are required to
prove maximum entropy. 

The beamsplitter addition is motivated by how an optical beamsplitter of transmissivity $\eta \in [0, 1]$ ``adds" the photon-number distributions of two classical mixtures of number states. It underlies the conjectural Entropy Photon Number Inequality (EPnI)~\cite{guha3,guha4}, which plays a role analogous to Shannon's EPI in understanding the capacity of Gaussian bosonic channels. The paper~\cite{qi} includes a more detailed history of the beamsplitter addition $\nadd{\eta}$. The key aim of the main part of the present paper however, is to give a self-contained presentation of the beamsplitter addition $\nadd{\eta}$ as a way of combining random variables supported on $\ZZ_+$,
in a way that is accessible to a purely classical audience.
Although some of our results may be known to the quantum information community, our hope is that this paper will stimulate future work by probabilists and classical information theorists on open problems---in particular, a proof of the conjectured EPI and entropic monotonicity under the beamsplitter addition~\cite{guha}.

We first define $\nadd{\eta}$ in the notation of \cite{guha}. That is:
\begin{definition} \label{def:beamsplit}
Given a random variable supported on $\ZZ_+$, define its
continuous counterpart $\com{X}$
(a circularly symmetric random variable supported on the complex plane $\CC$),
using a map $\TT$ with $\com{X} = \TT(X)$ and $X = \TT^{-1}(\com{X})$ with actions on mass
functions and densities given by \cite[Eq. (14),(15)]{guha}:
\begin{eqnarray}
p_{\com{X}}(\vc{r}) & = & \sum_{n=0}^\infty p_X[n] \frac{ e^{-|\vc{r}|^2} | \vc{r} |^{2n}}{n! \pi}, \label{eq:tdef} \\
p_X[n] & = & \frac{1}{\pi}
\int_{\CC} p_{\com{X}}(\vc{r}) \LL_n \left( | \vc{s} |^2 \right) \exp( \vc{r} \vc{s}^* - \vc{r}^* \vc{s}) d\vc{r} d\vc{s},\;\;\; \label{eq:tinvdef}
\end{eqnarray}
where $\LL_n$ denotes the $n^{\rm th}$ Laguerre polynomial.
As in \cite{guha}, for $0 \leq \eta \leq 1$,
we define the beamsplitter addition operation $\nadd{\eta}$ acting on random variables supported on $\ZZ_+$ by 
\begin{equation} \label{eq:nadddef}
X \nadd{\eta} Y = \TT^{-1} \left( \sqrt{ \eta}\, \TT( X ) + \sqrt{1-\eta}\, \TT(Y) \right),\end{equation}
where `$+$' on  the RHS of \eqref{eq:nadddef} denotes standard addition in $\CC$.
\end{definition}
The key contributions and  structure of the rest of the paper are as follows.
In Section \ref{sec:relation} we define two types of generating functions, for $X$ and $\com{X}$, and prove a new
relation between them in Theorem \ref{thm:hphi}.
 In Section \ref{sec:beamsplit}, we prove Theorem \ref{thm:transform}, which shows that
the generating function of $X \nadd{\eta} Y$ is a product of generating functions. In Section \ref{sec:debruijn}, we show 
that Theorem \ref{thm:transform} implies a heat equation (Theorem \ref{thm:pde}), which in turn gives  a
de Bruijn identity (Theorem \ref{thm:debruijn}). 
 In Section \ref{sec:logsum}, we state and prove a new bound on relative entropy under the action
of $\TT$. In Section \ref{sec:conc} we discuss some future work. The remainder of the main part of the paper contains proofs of
its main results. In an Appendix we review necessary ideas from quantum optics in order to show how the main results of the paper
may be understood in this context.
\section{Relation between generating functions} \label{sec:relation}
We consider two different kinds of generating functions, exponential and ordinary, recalling
that they are related by the Laplace transform (see Lemma \ref{lem:laplace}).
We write  $\ep \left( X \right)_{(m)} := \ep X(X-1) \ldots (X-m+1) = \ep X!/(X-m)!$ for the falling moment of a random variable
on $\ZZ_+$.

\begin{definition} \mbox{ }
\begin{enumerate}
\item
Given random variable $X$ with p.m.f. $p_X[m]$, $m \in \ZZ_+$, consider the sequence $ \ep \left( X \right)_{(m)}/m!$ and
 write:
\begin{enumerate} 
\item The ordinary generating function 
\begin{equation} \label{eq:discgen}
H_X(t) := \sum_{m=0}^\infty  t^m \left( \frac{\ep \left( X \right)_{(m)}}{m!} \right), \end{equation}
\item The   exponential generating function
\begin{equation} \label{eq:discgenexp}
\wth_X(t) := \sum_{m=0}^\infty \frac{t^m}{m!} \left( \frac{\ep \left( X \right)_{(m)}}{m!} \right). \end{equation}
\end{enumerate}
\item
For circularly symmetric $\com{X}$ supported on $\CC$ with density $p_{\com{X}}$,  
consider the sequence  $\frac{\ep | \com{X} |^{2m}}{m!}$ and 
write: 
\begin{enumerate}
\item The ordinary generating function
\begin{eqnarray} \label{eq:contgen}
\phi_{\com{X}}(t) & := & \sum_{m=0}^\infty t^m \left( \frac{\ep | \com{X} |^{2m}}{m!} \right) , 
\end{eqnarray}
\item The exponential generating function
\begin{equation} \label{eq:contgen2}
 \wtp_{\com{X}}(t) := \sum_{m=0}^\infty \frac{ t^m}{m!} \left( \frac{\ep | \com{X} |^{2m}}{m!} \right) .
\end{equation}
\end{enumerate}
\end{enumerate}
\end{definition}
Note that although \eqref{eq:discgen} and \eqref{eq:discgenexp} are defined as formal sums, in practice we focus on $t \leq 0$.
We first make the following claim, proved in Section \ref{sec:prooftransform}:
\begin{lemma} \label{lem:key}
For any random variable $X$ supported on $\ZZ_+$:
 \begin{enumerate}
\item $\wth_X(t) = \sum_{n=0}^\infty p_X[n] \LL_n(-t)$,
\item $p_X[m] = \int_0^\infty \exp(-s) \wth_X(-s) \LL_m(s) ds$. 
\end{enumerate}
\end{lemma}

\begin{example}  \label{ex:transforms}
For geometric $X$ with mean $\lambda$,
$\ep \left( X \right)_{(m)} =  m! \lambda^m$, so that $H_X(t)  = 1/(1- \lambda t)$
and $\wth_X(t)  = \exp( \lambda t)$.
Further  $\com{X}$ is circularly symmetric Gaussian with covariance matrix $(1+\lambda) I_2/2$,  the
$\ep | \com{X} |^{2m} = m! (1+\lambda)^m$, so $\phi_{\com{X}}(t) = 1/(1- (1+\lambda) t)$ and
$\wtp_{\com{X}}(t) = \exp( t(1+ \lambda))$.  
\end{example}
Example \ref{ex:transforms} illustrates the following result, which shows that  
$X$ and $\com{X}$ have a simple link at the level of their generating functions:
\begin{theorem} \label{thm:hphi} For $X$ and $\com{X}$ linked by the transforms
\eqref{eq:tdef} and \eqref{eq:tinvdef} of \cite{guha}, we can write
\begin{enumerate}
\item
\begin{equation} \wth_X(t) = \exp(-t) \wtp_{\com{X}}(t).\label{eq:expinv} \end{equation}
\item 
\begin{equation}
H_X(t)  =  \frac{1}{1+t} \phi_{\com{X}} \left( \frac{t}{1+t} \right),  \label{eq:hphi} 
\end{equation}
\end{enumerate}
 \end{theorem}
\begin{IEEEproof} See Section \ref{sec:prooftransform}. \end{IEEEproof}
This result relates the moments of $X$ and $\com{X}$. For brevity, from now on we write
$\lambda_W$ for the mean of any random variable $W$. Then,
for example:
\begin{corollary} \label{cor:Xmoments}
The real and imaginary parts of $\com{X} = (X_1,X_2)$ have covariance matrix $(1+\lambda_X) \Id_2/2$, where we 
write $\Id_d$ for the $d$-dimensional identity matrix.
\end{corollary}
\begin{IEEEproof}
We  simply differentiate \eqref{eq:hphi} with respect to $t$ and set $t=0$ to obtain:
$$ \lambda_X = H_X'(0) = \phi_{\com{X}}'(0) - \phi_{\com{X}}(0) = \ep | \com{X} |^2 - 1.$$
Since $\com{X}$ is circularly symmetric, it is proper (see \cite{rimoldi}), and
we know $\ep \com{X} = 0$. Further, this means that $(X_1,X_2)$ has a covariance matrix which is a multiple of the identity. 
We deduce that the diagonal entries must equal $(\lambda_X + 1)/2$.
\end{IEEEproof}
\section{Generating functions and beamsplitter addition} \label{sec:beamsplit}
We now state the relationship between the generating functions of $X$, $Y$ and $Z = X \nadd{\eta} Y$, 
 proved in Section \ref{sec:convproof}:
\begin{theorem} \label{thm:transform} Given independent random variables $X$ and $Y$ supported on $\ZZ_+$, the
 $Z := X \nadd{\eta} Y$ has generating functions  $\wth_Z$ and $H_Z$ satisfying:
\begin{enumerate}
\item
\begin{equation}
\label{eq:sub1} \wth_Z(t) = \wth_{X}( \eta t) \wth_{Y}((1-\eta) t). \end{equation}
\item
\begin{equation} \label{eq:sub2}
\frac{1}{s-1} H_Z \left( \frac{1}{s-1} \right) = \Lap{ \left( M_X^{(\eta)} \times M_Y^{(1-\eta)} \right) } \left( s  \right),
\end{equation}
where we define $M_X^{(\eta)}$ and $M_Y^{(1-\eta)}$ via the inverse Laplace transform $\Lapinv{\cdot}$ using
the fact that:
\begin{eqnarray}
\Lap{M_{\com{X}}^{(\eta)}}(s) & = & \frac{1}{s-\eta} H_X \left( \frac{ \eta}{s-\eta} \right), \label{eq:mxtran} \\
\Lap{M_{\com{Y}}^{(1-\eta)}}(s) & = & \frac{1}{s-(1-\eta)} H_Y \left( \frac{1- \eta}{s- (1-\eta)} \right). \;\;\; \label{eq:mytran}
\end{eqnarray}
\end{enumerate}
\end{theorem}
Direct calculation of the derivative of \eqref{eq:sub1}, as in Corollary \ref{cor:Xmoments}, allows us to deduce that
\begin{equation} \lambda_Z = \eta \lambda_X + (1-\eta) \lambda_Y. \label{eq:meanchange}
\end{equation}
\begin{example} \label{ex:preserved}
If $X$ is geometric with mean $\lambda_X$ and $Y$ is geometric with mean $\lambda_Y$, using the expressions from
Example \ref{ex:transforms} and \eqref{eq:meanchange}
then 
\begin{enumerate} 
\item The RHS of \eqref{eq:sub1} becomes 
$$ \exp( \eta \lambda_X t ) \exp( (1-\eta) \lambda_Y t) = \exp( \lambda_Z t),$$
 so that $Z = X \nadd{\eta} Y$ is geometric with mean $\lambda_Z$.
\item
The RHS of \eqref{eq:mxtran} is $1/(s- \eta(1+\lambda_X))$,
so the inverse Laplace transform gives that $M^{(\eta)}_{\com{X}}(t) = \exp( \eta(1+\lambda_X) t )$, with
$M^{(\eta)}_{\com{Y}}(t) = \exp( (1-\eta)(1+\lambda_Y) t )$. As we would expect, this means that 
$\left( M_X^{(\eta)} \times M_Y^{(1-\eta)} \right) = \exp ( (1 + \lambda_Z) t)$  This allows us to deduce that 
$$ \frac{1}{s-1} H_Z \left( \frac{1}{s-1} \right) = \frac{1}{s- (1 +\lambda_Z)},
$$
and changing variables via $u = 1/(s-1)$ we deduce that $H_Z(u) = 1/(1 - \lambda_Z u)$ as we would hope.
\end{enumerate}
\end{example}
\begin{remark} Theorem \ref{thm:transform} and Example \ref{ex:preserved} suggest the exponential generating
function $\wth_{X}$ is more amenable than the ordinary generating function $H_X$. We state both results for future
reference, but recommend the first formulation.
\end{remark}
\section{De Bruijn identity} \label{sec:debruijn}

Motivated by  \cite{stam}, we give a de Bruijn identity with respect to beamsplitter addition $\nadd{\eta}$.
The key result is the following discrete analogue of the heat equation, analogous to 
\cite[Corollary 4.2]{johnson21} in  the Poisson case:
\begin{theorem} \label{thm:pde}   For a given random variable $X$
consider $Z_\eta := X \nadd{\eta} Y$, where $Y$ is  geometric.
Writing $\lambda(\eta) = \lambda_{Z_\eta} = \eta \lambda_X+ (1- \eta) \lambda_Y$ we obtain
\begin{equation}\label{eq:pde} \frac{\partial}{\partial \eta} p_{Z_\eta}[n] := \Delta  
\left(   \frac{n}{\eta}  \left( p_{Z_\eta}[n-1] \lambda_Y - p_{Z_\eta}[n] (1+ \lambda_Y)    \right) \right), \end{equation}
where for any function $u$, we write $\Delta (u[n]) := u[n+1] - u[n]$.
\end{theorem}
\begin{IEEEproof} See Section \ref{sec:debrujnproof}. \end{IEEEproof}
\begin{definition} For a random variable $X$ with mass function $p_X$,
define two new p.m.f.s supported on $\ZZ_+$  by 
\begin{equation} \label{eq:plusminus}
p_{X}^+[n] = \frac{ (n+1) p_{X}[n+1]}{\lambda_X} \mbox{\;\;\; and \;\;\;}
p_{X}^-[n] = \frac{(n+1) p_{X}[n]}{1+\lambda_X}.\end{equation}
\end{definition}
This allows us to deduce the following de Bruijn identity, which is a specialization to number-diagonal states of the more
general de Bruijn identity proved by K\"{o}nig and Smith \cite{koenig}:
\begin{theorem} \label{thm:debruijn}
 Given $Z_\eta= X \nadd{\eta} Y$, where $Y$ is geometric with mean $\lambda_Y$, we can write $G_\eta$ for a geometric with mean $\lambda(\eta) = \eta \lambda_X+ (1-\eta) \lambda_Y$. Then
\begin{align*}
& \frac{\partial}{\partial \eta} D( Z_\eta \| G_\eta) \\
&  =  \frac{\lambda_Y (1+\lambda(\eta))}{\eta}  D( p_{Z_\eta}^- \| p_{Z_\eta}^+) +
\frac{(1+\lambda_Y) \lambda(\eta)}{\eta}
 D(p_{Z_\eta}^+ \| p_{Z_\eta}^-),
\end{align*}
where $p_{Z_\eta}^+$ and $p_{Z_\eta}^-$ are defined in terms of \eqref{eq:plusminus}.
\end{theorem}
If $X$ is itself geometric then so is $Z_\eta$, meaning that  $p_{Z_\eta}^+ = p_{Z_\eta}^-$ 
(a negative binomial mass function) and the
two relative entropy terms on the RHS of Theorem \ref{thm:debruijn} vanish as expected.

We focus on the case where $\lambda_X = \lambda_Y$, where 
the RHS of Theorem \ref{thm:debruijn} becomes a symmetrized relative entropy.
If $G$ is geometric
with $\ep G = \lambda_X$ then direct calculation gives that 
\begin{equation} \label{eq:entrelent} D(X \| G) = H(G) - H(X). \end{equation}
This allows us to deduce the following  log-Sobolev type inequality which
may be of independent interest:
\begin{corollary} \label{cor:logsob}
For any random variable $X$, 
if $G$ is geometric  with mean $\ep G = \lambda_X$ then:
\begin{equation} \label{eq:logsob}
D(X \| G) \leq \lambda_X (1+\lambda_X) \left( D( p_X^- \| p_X^+) + D( p_X^+ \| p_X^-) \right).\end{equation}
\end{corollary}
\begin{IEEEproof}
We consider $Z_\eta = X \nadd{\eta} Y$, where $Y$ is geometric with $\lambda_Y = \lambda_X$, and apply \cite[Theorem 5]{guha}, 
which tells us that $H( Z_\eta) \geq \eta H(X) + (1-\eta) H(Y)$. Combining this with \eqref{eq:entrelent}, we can write
$$ D(Z_\eta \| G) \leq \eta D(X \| G) + (1-\eta) D(Y \| G) = \eta D(X \| G),$$
or rearranging that (since $\eta \leq 1$)
$$ \frac{ D(Z_\eta \| G) - D(X \| G)}{\eta-1} \geq D(X \|G).$$
If $\eta \rightarrow 1$, the LHS becomes the derivative $\frac{\partial}{\partial \eta} D(Z_\eta \| G) |_{\eta=1}$, and
we deduce the result using Theorem \ref{thm:debruijn}.
\end{IEEEproof}
In the language of \cite{johnson11},
Theorem \ref{thm:pde} suggests that we can introduce a score
function $\rho_X$, defined as:
\begin{definition}
For a random variable $X$ with mass function $p_X$ and  mean $\lambda_X$,
define a score function
\begin{eqnarray}
 \rho_{X}[n]  & :=  & \frac{n p_{X}[n-1] \lambda_X}{p_{X}[n] (1+ \lambda_X)} - n,  \label{eq:plusscore} 
\end{eqnarray}
where we define $\rho_{X}[0] = 0$ to ensure  $\sum_{n=0}^\infty P_X[n] \rho_{X}[n] =  0$. 
We define two Fisher-type quantities  in terms of it:
\begin{eqnarray}
J^+(X) & :=  &  \sum_{n=1}^\infty \frac{p_X[n]}{n} \rho_X[n]^2, \\
J^-(X) & :=  &  \sum_{n=1}^\infty \frac{p_X[n]}{n + \rho_X[n]} \rho_X[n]^2. 
\end{eqnarray}
\end{definition}
Note that $\rho_X$ vanishes if and only if $X$ is geometric, so that $J^+(X)$ and $J^-(X)$ are $\geq 0$, with equality if and only
if $X$ is geometric. Further, if we choose $Y$ to be geometric with mean $\ep Y = \lambda_X$
then \eqref{eq:pde} becomes
$$  \frac{\partial}{\partial \eta} p_{Z_\eta}[n] := \frac{1+\lambda_X}{\eta} \Delta \left( p_{Z_\eta}[n] \rho_{Z_\eta}[n] \right).$$
where as before, we write $\Delta (u[n]) := u[n+1] - u[n]$.
Secondly, linearising the logarithm in Corollary \ref{cor:logsob}
 implies  a quadratic version of this result, in the spirit of \cite{bobkov3}:
\begin{corollary} \label{cor:logsobbob}
For any random variable $X$, 
if $G$ is geometric  with mean $\ep G = \lambda_X$ then:
$$ D(X \| G) \leq (1+\lambda_X) \left( J^+(X) + J^-(X) \right) .$$
\end{corollary}

\section{Log-sum inequality} \label{sec:logsum}
We  state one further result, which controls how the relative entropy behaves under the action of the map $\TT$.
\begin{theorem}\label{thm:logsum} Given two random variables $X$ and  $Y$, with $\com{X}= \TT(X)$ and 
$\com{Y} = \TT(Y)$ then:
$$ D( \com{X} \| \com{Y} ) \leq D( X \| Y).$$
\end{theorem}
\begin{IEEEproof} 
Writing $\phi_n(\vc{r}) := e^{-|\vc{r}|^2} | \vc{r} |^{2n}/(n! \pi)$ then, using \eqref{eq:tdef} 
and the log-sum inequality \cite[Theorem 2.7.1]{cover}, for any $\vc{r}$:
\begin{eqnarray*}
\lefteqn{ p_{\com{X}}(\vc{r}) \log \left( \frac{ p_{\com{X}}(\vc{r})}{ p_{\com{Y}}(\vc{r})} \right)} \\
 & = &
\left( \sum_{n=0}^\infty p_X[n] \phi_n(\vc{r}) \right)  \log \left( \frac{ \sum_{n=0}^\infty p_X[n] \phi_n(\vc{r})     }{  
\sum_{n=0}^\infty p_Y[n] \phi_n(\vc{r})} \right) \\
& \leq & 
\sum_{n=0}^\infty p_X[n] \phi_n(\vc{r}) \log \left( \frac{ p_X[n]}{p_Y[n]} \right).\end{eqnarray*}
Integrating over $\vc{r}$ we deduce the result since  $\int \phi_n(\vc{r}) d\vc{r} = 1$ for each $n$.
\end{IEEEproof}

\section{Conclusions} \label{sec:conc}
We have introduced a new purely classical representation for the beamsplitter addition operation $\nadd{\eta}$,
 with respect to the exponential generating function. We have deduced a heat equation, and recovered a purely classical proof of a special case of the de Bruijn identity of K\"{o}nig and Smith~\cite{koenig}.

In future work, we hope to use this formalism to prove discrete entropy results based around the geometric family,
analogous to the classical results proved for the continuous entropy based around the Gaussian family. In particular, we hope that our results can give insights into a proof of the conjectured discrete EPI under beamsplitter addition~\cite{guha}---a special case of the Entropy Photon Number Inequality~\cite{guha3,guha4}---as well as give insights into convergence to the geometric and the conjectured monotonic increase in entropy under repeated beamsplitter addition~\cite{guha}, analogous to the classical Central Limit Theorem convergence to Gaussians and `law of thin numbers'~\cite{johnson26} convergence to Poissons. 

\section{Proof of  Transform relation, Theorem \ref{thm:hphi}} \label{sec:prooftransform}

\begin{IEEEproof}[Proof of Lemma \ref{lem:key}] \mbox{ }
\begin{enumerate}
\item We reverse
the order of summation in \eqref{eq:discgenexp} to obtain
\begin{eqnarray}
\wth_X(t) & = & \sum_{m=0}^\infty  \frac{t^m}{m!} \sum_{n=m}^\infty \binom{n}{m} p_X[n] \nonumber \\
& = & \sum_{n=0}^\infty p_X[n] \sum_{m=0}^n \binom{n}{m} \frac{t^m}{m!} \nonumber \\
& = & \sum_{n=0}^\infty p_X[n] \LL_n(-t), \label{eq:proofrep}
\end{eqnarray}
since $\LL_n(-t) = \sum_{m=0}^n \binom{n}{m} \frac{t^m}{m!}$  (see \cite[Eq. (5.1.6)]{szego}).
\item  This result follows  on  integrating $\exp(-s) \LL_m(s)$ times both sides of \eqref{eq:proofrep} with $s = -t$,
using the orthogonality relation for Laguerre polynomials \cite[Eq. (5.1.1)]{szego},
$$\int_0^\infty \exp(-s) \LL_m(s) \LL_n(s) ds = \delta_{mn}.$$
\end{enumerate}
\end{IEEEproof}
Recall the standard result that the Laplace transform $L$ relates
 exponential and ordinary generating functions:
\begin{lemma} \label{lem:laplace}
Given a sequence $\vc{a} = (a_n)_{n=0,1, \ldots}$, if we write $H_{\vc{a}}(t) = \sum_{n=0}^\infty a_n t^n$
and $\wth_{\vc{a}}(t) = \sum_{n=0}^\infty a_n t^n/n!$,  then
\begin{equation} \label{eq:laplace} \Lap{\wth_{\vc{a}}} \left( u \right) = \frac{1}{u} H_{\vc{a}} \left( \frac{1}{u} \right).\end{equation}
\end{lemma}
\begin{IEEEproof}
This follows since 
\begin{eqnarray}
\Lap{\wth_{\vc{a}}}(u) & = & \sum_{n=0}^\infty \frac{a_n}{n!} \left( \int_0^{\infty} \exp( - s u) s^n ds \right) \nonumber \\
& = & \sum_{n=0}^\infty \frac{a_n}{n!} \left( \frac{n!}{u^{n+1}} \right) = \frac{1}{u} H_{\vc{a}} \left( \frac{1}{u} \right).
 \label{eq:laplace2} \end{eqnarray} \end{IEEEproof}

\begin{IEEEproof}[Proof of Theorem \ref{thm:hphi}]
We can express \eqref{eq:contgen2} in terms of the Bessel function $J_0$ (see \cite[Eq. (1.71.1)]{szego}),
which we substitute to obtain \eqref{eq:bessel} below. We obtain:
\begin{eqnarray}
\wtp_{\com{X}}(-t) & = & \int p_{\com{X}}(\vc{r}) \sum_{m=0}^\infty \frac{|\vc{r}|^{2m} (-t)^m}{m!^2} d\vc{r} \nonumber \\
& = & \int p_{\com{X}}(\vc{r}) J_0( 2 |\vc{r}| \sqrt{t}) d\vc{r}  \label{eq:bessel} \\
& = & \sum_{n=0}^\infty p_X[n] \int \frac{ e^{-|\vc{r}|^2} | \vc{r} |^{2n}}{n! \pi}
J_0( 2 |\vc{r}| \sqrt{t}) d\vc{r} \label{eq:subdens} \\
& = & 2 \sum_{n=0}^\infty p_X[n] \int_0^\infty \frac{ e^{-r^2} r^{2n+1}}{n!}
J_0( 2 r \sqrt{t}) dr \;\;\;\; \label{eq:changevar1} \\
& = &  \sum_{n=0}^\infty p_X[n] \int_0^\infty \frac{ e^{-u} u^{n}}{n!}
J_0( 2  \sqrt{u t}) du \label{eq:changevar2} \\
& = &  \sum_{n=0}^\infty p_X[n] \LL_n(t) \exp(-t), \label{eq:laguerre}
\end{eqnarray}
where \eqref{eq:subdens} follows by substituting \eqref{eq:tdef},  \eqref{eq:changevar1} follows by moving from Cartesian
coordinates $d\vc{r}$ to polar $r dr d\theta$, \eqref{eq:changevar2} uses $u =r^2$ and
\eqref{eq:laguerre} follows by \cite[Theorem 5.4.1]{szego}. The result follows by Lemma \ref{lem:key}.

Consider taking Laplace transforms of both sides of \eqref{eq:expinv}.
Using Lemma \ref{lem:laplace} the Laplace transform of the LHS
is 
\begin{equation} \label{eq:laplhs}
\Lap{ \wth_{X}}(u) = \frac{1}{u} H_X \left( \frac{1}{u} \right),
\end{equation}
Again by Lemma \ref{lem:laplace}, since the Laplace transform of $f(t) \exp(- t)$ is the Laplace 
transform of $f$ shifted by $1$,  the Laplace transform of the RHS   is
\begin{equation} \label{eq:laprhs}
 \Lap{ \wtp_{\com{X}}}(u+1) = \frac{1}{u+1} \phi_{\com{X}}  \left( \frac{1}{1+u} \right). \end{equation}
Equating \eqref{eq:laplhs} and \eqref{eq:laprhs}, the result follows taking $u = 1/t$.
\end{IEEEproof}
\section{Proof of convolution relation, Theorem \ref{thm:transform}} \label{sec:convproof}
We first state a result that shows how the moments of circularly symmetric random variables
on $\CC$ behave on convolution:
\begin{lemma} \label{lem:convolution}
Given independent circularly symmetric
 $\com{X}$ and $\com{Y}$ and writing $\com{Z} := \sqrt{\eta} \com{X} + 
\sqrt{1-\eta} \com{Y}$, we can write
\begin{equation} \label{eq:wtptran}
 \wtp_{\com{Z}}(t) = \wtp_{\com{X}} \left( \eta t \right) \wtp_{\com{Y}} \left( (1-\eta) t \right)
.\end{equation}
\end{lemma}
\begin{IEEEproof} Consider independent $\com{U}$ and $\com{V}$ and write $\com{W} = \com{U} + \com{V}$.
Then \cite[Eq. (3)]{goldman} gives
$$ \ep | \com{W} |^{2m} = \sum_{n=0}^m \binom{m}{n}^2 \ep | \com{U} |^{2n} \ep | \com{V} |^{2m-2n}.$$
Multiplying $t^m/(m!)^2$, and summing, 
we obtain that
$$ \wtp_{\com{W}}(t) = \wtp_{\com{U}}(t)  \wtp_{\com{V}}(t)$$
and the result follows by rescaling. 
\end{IEEEproof}
Putting all this together we obtain: 

\begin{IEEEproof}[Proof of Theorem \ref{thm:transform}]

1. This result follows directly on combining \eqref{eq:expinv} and \eqref{eq:wtptran}.

2. Relabelling $t = 1/(s-1)$ in \eqref{eq:hphi} and using Lemma \ref{lem:laplace}, for any random variable $U$ we obtain:
\begin{equation} \label{eq:touse} 
\Lap{\wtp_{\com{U}}}(s) = \frac{1}{s} \phi_{\com{U}} \left( \frac{1}{s} \right) = \frac{1}{s-1} H_{U} \left( \frac{1}{s-1} \right).
\end{equation}
Taking $U = Z$ in \eqref{eq:touse}, and using  Lemma \ref{lem:convolution}, we know that $\wtp_{\com{Z}} = M^{(\eta)}_{\com{X}} \times M^{(1-\eta)}_{\com{Y}}$
where $M_{\com{X}}^{(\eta)}(t) = \wtp_{\com{X}} \left( \eta t \right)$.
We can use the fact that if $F = \Lap{f}$ then
$ \Lap{f(a t)}(s) = \frac{1}{a} F \left( \frac{s}{a} \right)$ to deduce using \eqref{eq:touse} that
$$ \Lap{ M^{(\eta)}_{\com{X}} }(t) =   \frac{1}{\eta} \Lap{ \wtp_{\com{X}}} \left( \frac{t}{\eta} \right)
= \left. \frac{1}{\eta} \frac{1}{s-1} H_X \left( \frac{1}{s-1} \right) \right|_{s = t/\eta} $$
and \eqref{eq:mxtran} follows. A similar argument based on the fact that $M_{\com{Y}}^{(1-\eta)}(t) =  \wtp_{\com{Y}} \left( (1-\eta) t \right)$ allows us to deduce \eqref{eq:mytran}.
\end{IEEEproof}

\section{Proof of de Bruijn identity, Theorem \ref{thm:debruijn}} \label{sec:debrujnproof}
We first 
 prove  the heat equation Theorem \ref{thm:pde}:
\begin{IEEEproof}[Proof of Theorem \ref{thm:pde}]
By Lemma \ref{lem:key}.1) we can write 
\begin{equation} h(\eta;t) := \wth_{Z_{\eta}}(t) = \sum_{n=0}^\infty p_{Z_\eta}[n] \LL_n(-t).\label{eq:serieseta} 
\end{equation}
Using \eqref{eq:sub1} we  also write $h(\eta ; t) = \wth_X(\eta t) \exp( (1-\eta) \lambda_Y t)$ and 
observe that this satisfies
\begin{eqnarray} \label{eq:pde2}
\frac{\partial}{\partial \eta} h(\eta; t) = \frac{t}{\eta} \frac{\partial}{\partial t} h(\eta; t) - \frac{\lambda_Y t}{\eta}
h(\eta; t).
\end{eqnarray}
Hence, by differentiating \eqref{eq:serieseta} and using \eqref{eq:pde2}, we obtain
\begin{eqnarray}
\lefteqn{
\sum_{n=0}^\infty \frac{\partial}{\partial \eta} p_{Z_\eta}[n] \LL_n(-t)}  \nonumber
\\ 
& =&  \frac{t}{\eta} \frac{\partial}{\partial t} h(\eta; t) - \frac{\lambda_Y t}{\eta} h(\eta; t) \nonumber \\
& = & \frac{-t}{\eta} \sum_{n=0}^\infty p_{Z_\eta}[n] \biggl( \LL_n'(-t) + \lambda_Y \LL_n(-t) \biggr) \nonumber \\
& = & - \frac{\lambda_Y}{\eta} \sum_{n=0}^\infty p_{Z_\eta}[n] \biggl(n+1) \LL_{n+1}(-t)  - (n+1) \LL_n(-t)  \biggr) \nonumber \\
& & + \frac{1+\lambda_Y}{\eta} \sum_{n=0}^\infty p_{Z_\eta}[n] \biggl( n \LL_{n}(-t)  - n \LL_{n-1}(-t)   \biggr) 
\label{eq:heattocomp} \\
& = & \sum_{n=0}^\infty \Delta 
\left(   \frac{n}{\eta}  \left( p_{Z_\eta}[n-1] \lambda_Y - p_{Z_\eta}[n] (1+ \lambda_Y)    \right) \right)  \LL_n(-t) \nonumber 
\end{eqnarray}
here \eqref{eq:heattocomp} follows
using the fact that \cite[Eq. (5.1.14)]{szego} $z \LL_n'(z) = n \LL_n(z) - n \LL_{n-1}(z)$.
and using the
three-term relation for Laguerre polynomials, $- z \LL_n(z) = (n+1) \LL_{n+1}(z) - (2n+1) \LL_n(z) + n \LL_{n-1}(z)$ (see \cite[Eq. (5.1.10)]{szego}).
Comparing coefficients of $\LL_n(-t)$  we conclude the result holds.
\end{IEEEproof}

\begin{IEEEproof}[Proof of Theorem \ref{thm:debruijn}]
Using \eqref{eq:entrelent}, and writing $p_{\eta}$ for $p_{Z_\eta}$, we can express  $D( Z_{\eta} \| G_\eta)$ as
\begin{eqnarray*}
 \sum_{n=0}^\infty p_{\eta}[n] \log p_{\eta}[n]  - \lambda(\eta) \log \lambda(\eta) + (1+\lambda(\eta)) \log(1+ \lambda(\eta)).
\end{eqnarray*}
For any function $u[n]$, the
$\sum_{n=0}^\infty \Delta( u[n]) \log p_{\eta}[n] = \sum_{n=0}^\infty u[n+1] \log( p_{\eta}[n]/p_{\eta}[n+1])$, so
(assuming we can exchange the sum and derivative)
we obtain
\begin{eqnarray}
\lefteqn{\frac{\partial}{\partial \eta} D( Z_{\eta} \| G_\eta)} \nonumber \\
  &= & \sum_{n=0}^\infty \frac{\partial}{\partial \eta} p_{\eta}[n] \log p_{\eta}[n] -
\lambda'(\eta) \log \left( \frac{\lambda(\eta)}{1+ \lambda(\eta)} \right) \nonumber \\
&= &   \sum_{n=0}^\infty \frac{n+1}{\eta}  \left( p_{\eta}[n] \lambda_Y - p_{\eta}[n+1] (1+ \lambda_Y)    \right) \nonumber \\
&  & \;\;\;\;\;\; \times \log
\left( \frac{ p_{\eta}[n]/(1+\lambda(\eta))}{p_{\eta}[n+1]/\lambda(\eta)} \right) 
 \label{eq:move} 
\end{eqnarray}
where \eqref{eq:move} follows using \eqref{eq:meanchange}, and  adding factors of $(n+1)$ to the top and bottom of
the fraction to deduce the result.
\end{IEEEproof}

{\em Acknowledgements:} The authors thank the organizers of the 2016 Beyond i.i.d. in Information Theory workshop, where discussions leading up to this paper first took seed.


\appendix 

The goal of this Appendix is to provide alternative proofs of several results presented in this paper using tools from quantum optics. These alternative proofs are not required to follow any of our results in the main text. However, the reader familiar with the basic notations of quantum optics (briefly reviewed below) may find this discussion insightful. 

The Appendix is organized as follows. In Section~\ref{app:quantumreview}, we review the basics of quantum optics notation that we will need to describe our results. In Section~\ref{sec:transformrelations}, we provide simple alternative proofs of Lemma~\ref{lem:key}, Theorem~\ref{thm:hphi}, Theorem~\ref{thm:transform} and Lemma~\ref{lem:convolution} using quantum optics notation. In Section~\ref{sec:quantumdebruijn}, we show how the discrete variable analogue of the heat equation, Theorem~\ref{thm:pde} in the main text, can be derived as a special case of a quantum result from~\cite{koenig}. Finally, in Section~\ref{sec:logsum} we show that the log-sum identity, Theorem~\ref{thm:logsum} of the main text, can be interpreted as monotonicity of quantum relative entropy under the action of a trace-preserving completely-positive map.

\subsection{Brief review of quantum optics notation}\label{app:quantumreview}

\subsubsection{Bosonic modes, quantum states, characteristic functions}

The annihilation operator $a$ of a single mode of an electromagnetic field, and the creation operator $a^\dagger$, its Hermitian conjugate, act on the number states $\left\{|n\rangle\right\}$, $n \in {\mathbb Z}_+$ in the following way:
\begin{eqnarray}
a\,|n\rangle &=& \sqrt{n}\,|n-1\rangle, \,{\text{and}}\\
a^\dagger\,|n\rangle &=& \sqrt{n+1}\,|n+1\rangle,
\end{eqnarray}
with $\langle n | m \rangle = \delta_{nm}$. The number states form a complete orthonormal basis for the infinite-dimensional Hilbert space ${\cal H}$ of a single bosonic mode. A quantum state of a mode $\rho \in {\cal D}({\cal H})$, where ${\cal D}({\cal H})$ is the set of unit-trace positive Hermitian operators (called ``density operators") in ${\cal H}$.

A special class of states called {\em coherent states},
\begin{eqnarray}
|\alpha \rangle = e^{-|\alpha|^2/2} \sum_{n=0}^\infty \frac{\alpha^n}{\sqrt{n!}} |n\rangle
\end{eqnarray}
are eigenstates of $a$, i.e., $a\, |\alpha \rangle = \alpha \, |\alpha \rangle$ for all $\alpha \in {\mathbb C}$. Coherent states of a mode form an overcomplete basis, i.e.,
\begin{equation}
\int_{\mathbb C} \frac{|\alpha \rangle \langle \alpha |}{\pi}\, d^2\alpha = \sum_{n=0}^\infty\,|n\rangle \langle n| = {I}_\infty,
\label{eq:overcomplete}
\end{equation}
where ${I}_\infty$ is the identity operator in ${\cal H}$. The coherent state $|0\rangle$ is same as the number state $|0 \rangle$, known as the {\em vacuum state}.

The displacement operator is defined as
\begin{equation}
D(\alpha) = e^{-\alpha^* a + \alpha a^\dagger}, \,\alpha \in {\mathbb C}. 
\end{equation}
The reason it is called the displacement operator is that it displaces the vacuum state to a coherent state of a given complex amplitude, i.e., $|\alpha \rangle = D(\alpha)\,|0\rangle$, $\forall \alpha \in {\mathbb C}$. The matrix elements of $D(\alpha)$ in the number state basis are given by:
\begin{eqnarray}
\langle n | D(\alpha) | n \rangle &=& e^{-|\alpha|^2/2}\, {\cal L}_n\left(|\alpha|^2\right), \label{eq:D_diagonal}\\
\langle m | D(\alpha) | n \rangle &=& \sqrt{\frac{n!}{m!}} e^{-|\alpha|^2/2}\, \alpha^{m-n}\, {\cal L}_n^{(m-n)}\left(|\alpha|^2\right), \label{eq:D_offdiagonal}
\end{eqnarray}
where the latter holds for $m \ge n$. The following is a useful expression:
\begin{equation}
e^{\zeta a} |n\rangle = \sum_{m=0}^n \zeta^k \sqrt{\frac{1}{m!}\binom{n}{m}}\,|n-m\rangle .
\end{equation}

The average of an operator $O$ in state $\rho$, is $\langle O \rangle = {\rm tr}(\rho O)$. From hereon, we will start labeling the density operator by a subscript (e.g., $X$), where $X$ is the label of the mode whose state we are referring to. We define three characteristic functions of a density operator $\rho_X$, with $\zeta \in {\mathbb C}$, as follows:
\begin{eqnarray}
\chi_N^{\rho_X}(\zeta) &=& {\rm tr}\left(\rho_X\, e^{\zeta a_X^\dagger} e^{-\zeta^* a_X}\right) \,\, {\text{normal ordered}} \label{eq:def_normal}\\
\chi_A^{\rho_X}(\zeta) &=& {\rm tr}\left(\rho_X\, e^{-\zeta^* a_X} e^{\zeta a_X^\dagger}\right) \,\, {\text{anti-normal ordered}} \\
\chi_W^{\rho_X}(\zeta) &=& {\rm tr}\left(\rho_X\, e^{-\zeta^* a_X + \zeta a_X^\dagger}\right) \,\, {\text{Wigner}}
\end{eqnarray}
The Wigner characteristic function is the mean of the displacement operator $D_X(\zeta) = e^{-\zeta^* a_X + \zeta a_X^\dagger}$. It is simple to show that for any state $\rho_X$, its characteristic functions are related by:
\begin{eqnarray}
\chi_W^{\rho_X}(\zeta) &=& e^{|\zeta|^2/2}\,\chi_A^{\rho_X}(\zeta) \label{eq:CFequivalence_A}\\
&=& e^{-|\zeta|^2/2}\,\chi_N^{\rho_X}(\zeta). \label{eq:CFequivalence_N}
\end{eqnarray}
The Husimi function, popularly known as the $Q$-function, of a state $\rho_X$ is defined as:
\begin{equation}
Q_X(\alpha) = \frac{\langle \alpha | \rho_X | \alpha \rangle}{\pi}, \, \alpha \in {\mathbb C}.
\end{equation}
It readily follows from~\eqref{eq:overcomplete} that $Q_X(\alpha)$ is a proper probability distribution, i.e., $0 \le Q_X(\alpha) < 1/\pi$ and $\int_{\mathbb C} Q_X(\alpha)d^2\alpha = 1$ for any state $\rho_X$.

The $Q$-function and the anti-normal ordered characteristic function are related by a Fourier transform relationship:
\begin{eqnarray}
\chi_A^{\rho_X} &=& \int_{\mathbb C} Q_X(\alpha)\,e^{\zeta \alpha^* - \zeta^* \alpha}\,d^2\alpha, \,{\text{and}} \label{eq:QX_to_chi_A}\\
Q_X(\alpha) &=& \frac{1}{\pi^2} \int_{\mathbb C} \chi_A^{\rho_X}\,e^{-\zeta \alpha^* + \zeta^* \alpha}\,d^2\zeta .
\end{eqnarray}

The density operator can in turn be retrieved from the characteristic functions as follows:
\begin{eqnarray}
\rho_X &=& \int_{\mathbb C} \chi_A^{\rho_X}(\zeta) e^{-\zeta a_X^\dagger} e^{\zeta^* a_X} \frac{d^2\zeta}{\pi} \label{eq:inverse_A}\\
&=& \int_{\mathbb C} \chi_W^{\rho_X}(\zeta) D^\dagger_X(\zeta)\frac{d^2\zeta}{\pi} \label{eq:inverse_W}\\
&=& \int_{\mathbb C} e^{-|\zeta|^2/2}\chi_N^{\rho_X}(\zeta) D^\dagger_X(\zeta) \frac{d^2\zeta}{\pi}.\label{eq:inverse_N}
\end{eqnarray}

It follows from Eqs.~\eqref{eq:D_diagonal}, ~\eqref{eq:CFequivalence_N} and the fact that ${\cal L}(-t) = \sum_{m=0}^n \binom{n}{m}\frac{t^m}{m!}$, that the normal ordered characteristic function of the number state $|n\rangle$, i.e., $\rho_X = |n\rangle \langle n |$ is given by,
\begin{eqnarray}
\chi_N^{|n\rangle \langle n |}(\zeta) &=& {\cal L}_n\left(|\zeta|^2\right) \label{eq:chiN_Laguerre}\\
&=& \sum_{m=0}^n \binom{n}{m} \frac{\left(-|\zeta|^2\right)^m}{m!}.\label{eq:chiN_numberstate}
\end{eqnarray}

\subsubsection{Input output relationship of a beamsplitter}\label{sec:beamsplitterinputoutput}

The Heisenberg description of the input-output relationship of a beamsplitter of transmissivity $\eta$ is given by:
\begin{equation}
a_Z = \sqrt{\eta}\,a_X + \sqrt{1-\eta}\,a_Y,
\label{eq:beamsplitter}
\end{equation}
where $a_X$, $a_Y$ and $a_Z$ are annihilation operators of two input modes (X and Y) and an output mode (Z), respectively. We take $\rho_X$, $\rho_Y$ and $\rho_Z$ to be the density operators of the quantum states of the respective modes. Let us also assume the input states are statistically independent (viz., in a product state), i.e., $\rho_{XY} = \rho_X \otimes \rho_Y$. 

It readily follows from~\eqref{eq:beamsplitter}, and the fact that $\rho_X$ and $\rho_Y$ are statistically independent, that the characteristic functions undergo a scaled multiplication under the beamsplitter mixing. 
\begin{eqnarray}
\chi_N^{\rho_Z}(\zeta) = \chi_N^{\rho_X}(\sqrt{\eta}\,\zeta)\, \chi_N^{\rho_Y}(\sqrt{1-\eta}\,\zeta), \label{eq:chiZ_prod_N}\\
\chi_A^{\rho_Z}(\zeta) = \chi_A^{\rho_X}(\sqrt{\eta}\,\zeta)\, \chi_A^{\rho_Y}(\sqrt{1-\eta}\,\zeta), \label{eq:chiZ_prod_A}\\
\chi_W^{\rho_Z}(\zeta) = \chi_W^{\rho_X}(\sqrt{\eta}\,\zeta)\, \chi_W^{\rho_Y}(\sqrt{1-\eta}\,\zeta).\label{eq:chiZ_prod_W}
\end{eqnarray}
This is reminiscent of how the probability distribution functions of statistically-independent random variables $X$ and $Y$ undergo a scaled convolution under the operation $Z = \sqrt{\eta} X + \sqrt{1-\eta} Y$, and that the Fourier transforms of the pdfs undergo a scaled multiplication.

\subsubsection{Number diagonal states}\label{sec:diagonal}

Let us consider a state $\rho_X = \sum_{n=0}^\infty p_X[n]|n\rangle \langle n|$, which is diagonal in the number basis. The unit trace condition of a density operator, ${\rm tr}(\rho_X) = 1$, implies that $\left\{ p_X[n] \right\}$, $n \in {\mathbb Z}_+$ is a proper p.m.f., where ${\mathbb Z}_+$ is the set of non-negative integers. We interpret $p_X[n]$ as the p.m.f. of a discrete-valued random variable $X$ that takes values in ${\mathbb Z}_+$. The Q-function $Q_X(\alpha)$ is circularly symmetric (i.e., only a function of the magnitude $|\alpha|$) if and only if $\rho_X$ is number diagonal [CITE]. Interpreting $Q_X(\alpha)$ as the p.d.f. of a circularly-symmetric continuous-valued random variable ${\boldsymbol X}_c$, simple algebra leads us to the following $1$-to-$1$ relationship between the p.d.f. of ${\boldsymbol X}_c$ and the p.m.f. of $X$:
\begin{eqnarray}
p_{\com{X}}(\vc{r}) & = & \sum_{n=0}^\infty p_X[n] \frac{ e^{-|\vc{r}|^2} | \vc{r} |^{2n}}{n! \pi}, \label{eq:tdef_app} \,{\text{and}}\\
p_X[n] & = & \frac{1}{\pi}
\int_{\CC} p_{\com{X}}(\vc{r}) \LL_n \left( | \vc{s} |^2 \right) \exp( \vc{r} \vc{s}^* - \vc{r}^* \vc{s}) d\vc{r} d\vc{s},\;\;\; \label{eq:tinvdef_app}
\end{eqnarray}
which was introduced in~\cite{guha} as the pair of maps, $\com{X} = \TT(X)$ and $X = \TT^{-1}(\com{X})$. 

If two number diagonal states $\rho_X = \sum_{n=0}^\infty p_X[n]|n\rangle \langle n|$ and $\rho_Y = \sum_{n=0}^\infty p_Y[n]|n\rangle \langle n|$ interfere on a beamsplitter of transmissivity $\eta$~\eqref{eq:beamsplitter}, the output state is also number diagonal, $\rho_Z = \sum_{n=0}^\infty p_Z[n]|n\rangle \langle n|$. We define ${\boldsymbol Y}_c$ and ${\boldsymbol Z}_c$ analogously from using~\eqref{eq:tdef}. Since the $Q$-functions $p_{{\boldsymbol X}_c}({\boldsymbol r})$, $p_{{\boldsymbol Y}_c}({\boldsymbol r})$ and $p_{{\boldsymbol Z}_c}({\boldsymbol r})$ are Fourier inverses of the respective antinormal-ordered characteristic functions $\chi_A^{\rho_X}(\zeta)$, $\chi_A^{\rho_Y}(\zeta)$ and $\chi_A^{\rho_Z}(\zeta)$, given the characteristic functions multiply as in~\eqref{eq:chiZ_prod_A}, we conclude that ${\boldsymbol Z}_c = \sqrt{\eta}{\boldsymbol X}_c + \sqrt{1-\eta}\,{\boldsymbol Y}_c$. As in \cite{guha}, for $0 \leq \eta \leq 1$, we define the beamsplitter addition operation $\nadd{\eta}$ acting on random variables supported on $\ZZ_+$ by 
\begin{equation} \label{eq:nadddef_app}
Z = X \nadd{\eta} Y = \TT^{-1} \left( \sqrt{ \eta} \TT( X ) + \sqrt{1-\eta} \TT(Y) \right),\end{equation}
where the `$+$' on  the RHS of \eqref{eq:nadddef_app} is standard addition in $\CC$.

\subsection{Transform relations involving discrete and continuous exponential generating functions}\label{sec:transformrelations}

Let us consider a discrete-valued random variable $X \in {\mathbb Z}_+$, a corresponding circularly-symmetric continuous-valued random variable $\com{X} \in {\mathbb C}$, and a number-diagonal state $\rho_X$ with photon number distribution $p_X$, as described in Section~\ref{sec:diagonal}. Let us evaluate $\chi_N^{\rho_X}(\zeta)$ using the definition~\eqref{eq:def_normal} and the normal ordered characteristic function of a number state~\eqref{eq:chiN_numberstate}:
\begin{eqnarray}
\chi_N^{\rho_X}(\zeta) &=& {\rm tr}\left(\rho_X\, e^{\zeta a_X^\dagger} e^{-\zeta^* a_X}\right) \\
&=& \sum_{n=0}^\infty p_X[n] \chi_N^{|n\rangle \langle n |}(\zeta) \nonumber \\
&=& \sum_{n=0}^\infty p_X[n]  \sum_{m=0}^n \frac{n!}{(m!)^2 (n-m)!}\left(-|\zeta|^2\right)^m \nonumber \\
&=& \sum_{m=0}^\infty \frac{(-|\zeta|^2)^m}{(m!)^2} \sum_{n=m}^\infty p_X[n]\frac{n!}{(n-m)!}.
\end{eqnarray}
Therefore, $\chi_N^{\rho_X}(\zeta)$ is nothing but the exponential generating function of $X$ (see \eqref{eq:discgenexp})
\begin{equation} \label{eq:discgenexpQ}
\wth_X(t) = \sum_{m=0}^\infty \frac{t^m}{m!} \frac{\ep \left( X \right)_{(m)}}{m!}, 
\end{equation}
evaluated at $t = -|\zeta|^2$. Here we wrote $\ep \left( X \right)_{(m)} := \ep X(X-1) \ldots (X-m+1) = \ep X!/(X-m)!$ for the falling moment of a random variable on $\ZZ_+$. In other words,
\begin{equation}
\chi_N^{\rho_X}(\zeta) = \wth_X(-|\zeta|^2).
\label{eq:chiN_Htilde}
\end{equation}
Using Eqs.~\eqref{eq:chiN_Laguerre} and~\eqref{eq:chiN_Htilde}, we get
\begin{equation}
\wth_X(-|\zeta|^2) = \sum_{n=0}^\infty p_X[n] {\cal L}_n\left(|\zeta|^2\right).
\end{equation}
Next, we evaluate the number-basis diagonal elements of $\rho_X$ using Eq.~\eqref{eq:inverse_N}, and use~\eqref{eq:D_diagonal} to obtain:
\begin{eqnarray}
p_X[n] &=& \langle n | \rho_X | n \rangle \nonumber \\
&=& \int_{\mathbb C} e^{-|\zeta|^2/2} \chi_N^{\rho_X}(\zeta) e^{-|\zeta|^2/2} {\cal L}_n\left(|\zeta|^2\right)\frac{d^2\zeta}{\pi} \nonumber \\
&=& \int_{0}^\infty e^{-u}\wth_X(-u){\cal L}_n(u) du,
\end{eqnarray}
where in the last step we expressed $\zeta = re^{i\theta}$, used the substitution $|\zeta|^2 = r^2 = u$, $d^2\zeta = 2dr\,d\theta$, integrated over $\theta \in (0, 2\pi]$, and used the relationship~\eqref{eq:chiN_Htilde}.

Hence we have the following (Lemma~\ref{lem:key} of the main text):
\begin{lemma} \label{lem:key_app}
For any random variable $X$:
 \begin{enumerate}
\item $\wth_X(t) = \sum_{n=0}^\infty p_X[n] \LL_n(-t)$, for $t \in (-\infty, 0]$, and
\item $p_X[m] = \int_0^\infty \exp(-u) \wth_X(-u) \LL_m(u) du$.
\end{enumerate}
\end{lemma}

Now, let us consider the exponential generating function of \eqref{eq:contgen2}, that is
\begin{equation} \label{eq:contgen2Q}
 \wtp_{\com{X}}(t) = \sum_{m=0}^\infty \frac{ t^m}{m!} \left( \frac{\ep | \com{X} |^{2m}}{m!} \right).
\end{equation}
It can be shown by evaluating Eq.~\eqref{eq:QX_to_chi_A} for circularly-symmetric $Q_X(\alpha)$, that
\begin{equation}
\chi_A^{\rho_X}(\zeta) = \wtp_X(-|\zeta|^2).
\end{equation}
Using $\chi_A^{\rho_X}(\zeta) = e^{-|\zeta|^2}\chi_N^{\rho_X}(\zeta)$ (see~\eqref{eq:CFequivalence_A} and~\eqref{eq:CFequivalence_N}), we get the following theorem (Theorem~\ref{thm:hphi} of the main text):
\begin{theorem} \label{prop:hphi} For $X$ and $\com{X}$ linked by the transforms
\eqref{eq:tdef} and \eqref{eq:tinvdef} of \cite{guha}, we can write
\begin{eqnarray}
H_X(t)  &=&  \frac{1}{1+t} \phi_{\com{X}} \left( \frac{t}{1+t} \right), \,{\text{and}} \label{eq:hphiQ} \\
\wth_X(t) &=& \exp(-t) \wtp_{\com{X}}(t).\label{eq:expinvQ} \end{eqnarray}
 \end{theorem}

Finally, since we interpreted $\wtp_{\com{X}}(t)$ and $\wth_X(t)$ as characteristic functions, using Eqs.~\eqref{eq:chiZ_prod_N} and~\eqref{eq:chiZ_prod_A}, we obtain the following (Theorem~\ref{thm:transform} and Lemma~\ref{lem:convolution} of the main text, respectively):
\begin{eqnarray}
\wth_Z(t) &=& \wth_{X}( \eta t) \wth_{Y}((1-\eta) t), \, {\text{and}}\\
\wtp_{\com{Z}}(t) &=& \wtp_{\com{X}} \left( \eta t \right) \wtp_{\com{Y}} \left( (1-\eta) t \right).
\end{eqnarray}

\subsection{The discrete variable heat equation}\label{sec:quantumdebruijn}

A bosonic Gaussian channel (BGC) can be characterized by its action on the characteristic function of the input state. In particular, for one-mode gauge-covariant BGCs, the transformation of the input state $\rho_{\rm in}$ to the output state $\rho_{\rm out}$, expressed in terms of their characteristic functions assumes the following form (we are using the Wigner characteristic functions below):
	\begin{align}
	\label{eq:characteristic}
	\chi_{\rho_{\rm out}}(\xi) = \chi_{\rho_{\rm in}}(\sqrt{\tau}\,\xi)\exp[-y|\xi|^2/2],
	\end{align}
	where $\tau >0$ is the loss or gain parameter and $y$ parametrizes added (Gaussian) noise. The BGC is a valid TPCP map if $y\geq|\tau-1|$. For a lossy bosonic channel of transmissivity $\eta \in (0, 1)$ and added thermal noise of mean photon number $N$, $\tau = \eta$ and $y = (1-\eta)(2N+1)$. For a phase-insensitive bosonic amplifier of gain $\kappa > 1$ and added thermal noise of mean photon number $N$, $\tau = \kappa$ and $y = (\kappa - 1)(2N+1)$. For a unit-gain additive thermal noise channel with photon-number-unit noise variance $N$, $\tau = 1$ and $y = 2N$. All single-mode gauge-covariant BGCs posses a semi-group structure, and consequently that a gauge-covariant single-mode BGC can be represented as a one-parameter linear TPCP map~\cite{gio10},
	\begin{align}
	\rho(t)=\Phi_t(\rho) = e^{t\L}\rho~,
	\end{align}
	with manifest semi-group structure 
	\begin{align}
	e^{(t+t')\L} = e^{t'\L}e^{t\L}=e^{t\L}e^{t'\L}~.
	\end{align}
	Here $\L$ is a Lindblad operator that generates the dynamics of a gauge-covariant BGC, and $t$ can be viewed as a time parameter corresponding to a continuous action of the channel on the input state $\rho(0)$, resulting in the final state $\rho(t)$ at time $t > 0$. With that interpretation, the equation of motion for $\rho(t)$, under the action of the channel, is given by
	\begin{align}
	\frac{d\rho(t)}{dt} = \L\rho(t)~.
	\label{eq:quantumdeB}
	\end{align}
	For Gauge-covariant BGCs, the Lindblad operator is given by
	\begin{align}
	\L = \gamma_+ \L_+ + \gamma_-\L_-~,
	\end{align}
where
	\begin{align}
	\L_+(\rho) &= {\hat a}^\dagger\rho {\hat a} -\frac{1}{2}{\hat a}{\hat a}^\dagger\rho -\frac{1}{2} \rho {\hat a}{\hat a}^\dagger ~,\\
	\L_-(\rho) &= {\hat a}\rho {\hat a}^\dagger -\frac{1}{2}{\hat a}^\dagger {\hat a} \rho -\frac{1}{2} \rho {\hat a}^\dagger {\hat a}~.
	\end{align}
	For a lossy channel with additive thermal noise, $\gamma_+ = N,\gamma_- = N+1$, where $N$ is the mean photon number of the thermal state. The channel transmissivity $\eta = e^{-t}$. For a phase-insensitive noisy amplifier channel, $\gamma_+ = N+1,\gamma_- = N$, and the amplifier gain $\kappa = e^{t}$. Finally, for an additive Gaussian noise channel, $\gamma_+=\gamma_- = 1$ with $N=t$.

When $\rho_X$ is number diagonal, and $\rho_Y$ a number diagonal with geometric $p_Y[n]$ (thermal state) of mean $\lambda_Y$, it is simple to verify that ~\eqref{eq:quantumdeB} reduces to the following (Theorem~\ref{thm:pde} of the main text):
\begin{theorem} \label{thm:pdeQ}   For a given random variable $X$
consider $Z_\eta := X \nadd{\eta} Y$, where $Y$ is a geometric with mean $\lambda_Y$.
Writing $\lambda(\eta) = \eta \lambda_X+ (1- \eta) \lambda_Y$ for the mean of $Z_\eta$ we obtain
\begin{equation}\label{eq:pdeQ} \frac{\partial}{\partial \eta} p_{Z_\eta}[n] := \Delta  
\left(   \frac{n}{\eta}  \left( p_{Z_\eta}[n-1] \lambda_Y - p_{Z_\eta}[n] (1+ \lambda_Y)    \right) \right), \end{equation}
where for any function $u$, we write $\Delta (u[n]) := u[n+1] - u[n]$.
\end{theorem}

\subsection{Log-sum inequality as monotonicity of quantum relative entropy under the action of a TPCP map}\label{sec:logsumQ}

Given two quantum states $\rho, \sigma \in {\cal D}({\cal H}_1)$, and a quantum channel---a trace-preserving completely positive (TPCP) map---$T: {\cal D}({\cal H}_1) \to {\cal D}({\cal H}_2)$, we have that~\cite{lindblad}:
\begin{equation}
D(\rho || \sigma) \ge D\left(T(\rho)||T(\sigma)\right),
\label{eq:monotonicity_QRE}
\end{equation}
where $D(\rho || \sigma) = {\rm Tr}\left\{\rho\left(\log \rho - \log \sigma\right)\right\}$ is the quantum relative entropy between the states $\rho$ and $\sigma$. If $\rho$ and $\sigma$ are simultaneously diagonalizable, i.e., $\rho = \sum_{k}p[k] |k\rangle \langle k|$ and $\sigma = \sum_{k}q[k] |k\rangle \langle k|$ for some complete orthonormal basis $\left\{|k\rangle\right\}$ in ${\cal H}_1$, then the quantum relative entropy equals the classical relative entropy between the probability distributions in the common spectral basis, i.e., $D(\rho || \sigma) = D(p || q)$. Eq.~\eqref{eq:monotonicity_QRE} states that the action of a channel cannot increase the quantum relative entropy between a pair of input states. 

Any quantum measurement can be interpreted as a (measure-and-prepare) quantum channel. Therefore, if a measurement, described by POVM operators $\left\{E_k\right\}$, act on states $\rho$ and $\sigma$ to induce probability distributions $p[k]$ and $q[k]$ respectively, i.e., ${\rm tr}(\rho E_k) = p[k]$ and ${\rm tr}(\sigma E_k) = q[k]$,~\eqref{eq:monotonicity_QRE} reduces to $D(\rho || \sigma) \ge D(p || q)$. 

Consider number diagonal states $\rho_X = \sum_{n=0}^\infty p_X[n]|n\rangle \langle n|$ and $\rho_Y = \sum_{n=0}^\infty p_Y[n]|n\rangle \langle n|$. Since $\rho_X$ and $\rho_Y$ are both diagonal in the number basis, $D(\rho_X || \rho_Y) = D(X||Y)$, where the RHS is the classical relative entropy between random variables $X$ and $Y$ whose p.m.f.s are $p_X$ and $p_Y$ respectively. Let us consider measuring $\rho_X$ and $\rho_Y$ in the coherent-state (overcomplete) basis $\left\{\left(|\alpha\rangle \langle \alpha |\right)/\pi\right\}, \alpha \in {\mathbb C}$. See~\eqref{eq:overcomplete}. The measurement by definition induces the continuous circularly-symmetric distributions $p_{{\boldsymbol X}_c}$ and $p_{{\boldsymbol Y}_c}$, the respective Husimi functions of $\rho_X$ and $\rho_Y$. Therefore, the statement of Theorem~\ref{thm:logsum} follows.

\end{document}